\documentclass{aa}
\usepackage[dvips]{graphicx}
\def\lesssim{\mathrel{\hbox{\rlap{\hbox{\lower4pt\hbox{$\sim$}}}\hbox{$<$}}}}
\def\gtrsim{\mathrel{\hbox{\rlap{\hbox{\lower4pt\hbox{$\sim$}}}\hbox{$>$}}}}
 
\begin{document}

\thesaurus{11.05.2; 11.06.1; 11.07.1}
\title{New Clues on the Nature of Extremely Red Galaxies \thanks{Based 
on observations made at the European Southern Observatory, La Silla and
Paranal, Chile}}

%\subtitle{}
 
\author{A. Cimatti \inst{1}, E. Daddi \inst{2}, S. di Serego Alighieri 
\inst{1}, L. Pozzetti \inst{1}, F. Mannucci \inst{3}, A. Renzini \inst{4}, 
E. Oliva \inst{1}, G. Zamorani \inst{5}, P. Andreani \inst{6}, 
H.J.A. R\"ottgering \inst{7}}
\offprints{A. Cimatti}

\institute
{Osservatorio Astrofisico di Arcetri, Largo Fermi 5, I-50125,
Firenze, Italy\\e-mail: cimatti@arcetri.astro.it
\and Dipartimento di Astronomia, Universit\`a di Firenze, Largo Fermi 5,
I-50125, Firenze, Italy
\and CAISMI-CNR, Largo Fermi 5, I-50125, Firenze, Italy
\and European Southern Observatory, Karl-Schwarzschild-Str. 2, D-85748 Garching,
Germany
\and Osservatorio Astronomico di Bologna, via Ranzani 1, I-40127 Bologna, Italy
\and Osservatorio Astronomico di Padova, vicolo dell' Osservatorio 5, I-35122,
Padova, Italy
\and Sterrewacht Leiden, Sterrewacht, Postbus 9513, Leiden 2300 RA, The
Netherlands
}

\date{Received date; accepted date}
\titlerunning{Extremely Red Galaxies}
\authorrunning{A. Cimatti et al.}

\maketitle

\begin{abstract}

We present near-infrared VLT-UT1+ISAAC spectroscopy of a sample
of 9 extremely red galaxies (ERGs) with $R-K>5$ and $K<19.0$. 
Neither strong emission lines ($F_{{\rm lim}}<$1-5$\times10^{-16}$ 
erg s$^{-1}$cm$^{-2}$) nor continuum breaks are detected. From 
near-infrared spectrophotometry, complemented with broad-band 
optical and near-IR photometry, we estimate ``spectro-photometric'' 
redshifts to be in the range of $0.8 \lesssim z_{{\rm sphot}} \lesssim 1.8$. 
We derive upper limits on the star formation rates in range of $SFR < 
6-30$$h_{50}^{-2} $M$_{\odot}$yr$^{-1}$. Two of the observed ERGs 
are dusty starburst candidates because they require strong dust reddening 
to reproduce their global spectral energy distributions. The other ERGs 
are consistent with being dustless old passively evolved spheroidals at 
$z \gtrsim 0.8$. We discuss the general implications of our findings in 
relation with the problem of the formation of early type galaxies. 


\end{abstract}

\section{Introduction}

The combination of optical and near-infrared surveys has led to the 
discovery of a population of galaxies with extremely red colors (
hereafter called ERGs, sometimes also called extremely red objects, 
EROs, e.g. Elston, Rieke \& Rieke 1988; McCarthy et al. 1992; Hu \& 
Ridgway 1994). However, depending on the depth of the surveys and on the 
adopted filters, different color criteria have been used to classify a galaxy 
as ``extremely red''. Here we adopt a color threshold $R-K>5$, which 
corresponds to the observed colors of cluster and field ellipticals at 
$z \gtrsim 1$ (e.g. Spinrad et al. 1997; Rosati et al. 1999) and to old 
passively evolving populations at similarly high redshifts according to 
the Bruzual \& Charlot (1998) spectral synthesis models. According to 
such models, a color selection threshold $I-K>4$ is equivalent to 
$R-K>5$ in the selection of elliptical galaxy candidates at $z \gtrsim 
1$ (see for instance Barger et al. 1999).

ERGs are ubiquitous objects, being found in ``empty'' sky fields 
(e.g. Thompson et al. 1999), in the vicinity of high-$z$ AGN (McCarthy et 
al. 1992; Hu \& Ridgway 1994) and as counterparts of faint X-ray (Newsam 
et al. 1997) and radio sources (Spinrad et al. 1997). Very little 
is presently known about ERGs mainly because their faintness hampers 
spectroscopic observations. Their colors are 
consistent with two radically different scenarios: (1) ERGs are evolved 
spheroidals at $z \gtrsim 1$, their colors being due to the lack of 
star formation and to the strong K-correction, (2) ERGs are high-$z$ active 
or starburst galaxies heavily reddened by dust extinction. Such
scenarios are relevant both to understand the formation and the 
evolution of elliptical galaxies and to investigate the existence 
of a population of dusty galaxies or AGN strongly reddened by dust 
extinction. Recent observations suggested the existence of at least 
two populations of objects contributing to the ERG population.
On the one hand, near-IR spectroscopy and submillimeter observations 
showed that HR10 ($I-K\sim6.5$, Hu \& Ridgway 1994) is a dusty starburst 
galaxy at $z=1.44$ (Graham \& Dey 1996; Cimatti et al. 1998; Dey et 
al. 1999). On the other hand, further observations suggested that other 
ERGs are likely to be high-$z$ ellipticals (e.g. Spinrad et al. 1997; 
Stiavelli et al. 1999; Soifer et al. 1999). 

In order to assess the relative contribution of different galaxy types
to the overall ERG population 
and their respective role in galaxy evolution, we started a project based 
on optical and near-IR observations made with the ESO and {\it Hubble Space
Telescope} (HST) telescopes. We present here the first results of near-IR 
spectroscopy of a sample of ERGs. $H_0=50$ kms$^{-1}$ Mpc$^{-1}$ and 
$\Omega_0=1.0$ are assumed throughout the paper. 

\section{Sample selection, observations and analysis}

We surveyed about 95 arcmin$^{2}$ in high-$z$ AGN fields and in ``empty'' 
sky fields (Cimatti et al.; Daddi et al.; Pozzetti et al., in preparation), 
and we selected a complete sample of ERGs with $K<19.0$. Since homogeneous 
optical imaging was not always available to us for all the observed fields, 
we adopted a color selection threshold $R-K>5$ or $I-K>4$ depending on the 
available optical photometry. We do not 
expect contamination by low mass stars or brown dwarfs (e.g. Cuby et al. 
1999) because our color-selected ERGs turned out to have non-stellar 
morphologies. Here we present the first results for an incomplete ``pilot'' 
subsample that was selected according to the observability of the targets 
during the allocated VLT nights. 

$ZJHKs$-band imaging was done on 27-30 March 1999 with the imager-spectrograph 
SOFI (Moorwood, Cuby \& Lidman 1998) at the ESO 3.5m New Technology Telescope 
(NTT). In addition, $H$ and $Ks$ images of the J2027-217 field (with 10 and 
14 minutes exposure respectively) were obtained with the near-IR imager-
spectrograph ISAAC (Moorwood et al. 1999) at the ESO Very Large Telescope 
(VLT) UT1 ({\it Antu}) on 28 April 1999. Both SOFI and ISAAC are equipped 
with 1024$^{2}$ Rockwell detectors with scales of 0.$^{\prime\prime}$292/pixel 
and 0.$^{\prime\prime}$147/ pixel respectively. The observations were made 
during photometric conditions and with the seeing 0.7$^{\prime\prime}$-1.0$
^{\prime\prime}$ and 0.4$^{\prime\prime}$-0.5$^{\prime\prime}$ for SOFI and 
ISAAC respectively. During the observations the telescope was moved
between exposures according to a random pattern of offsets. The total 
integration times of the SOFI images were 30, 20, 15, 25 minutes in the 
$ZJHKs$ bands respectively. The $ZJHK$ photometric calibration was achieved 
with the standard 
stars of Persson et al. (1998) and Feige 56 and LTT 3864 (Hamuy et al. 1994). 
The night-to-night scatter of the zero points was about 0.02 magnitudes. 
Basic information on the optical photometry can be found in the captions
of Fig. 1 and 2, and more details will be given in forthcoming papers.

Photometry was carried out with SExtractor (Bertin \& Arnouts 
1996) using a 3$^{\prime\prime}$ diameter aperture and taking into
account aperture losses due to the different seeing conditions. The 
magnitudes were dereddened for Galactic extinction using the maps of
Burnstein \& Heiles (1982). The results in sections 3.2 and 3.3 do not 
change if dereddening is made with the Schlegel, Finkbeiner \& Davis
(1998) dust maps.

$JHK$-band spectroscopy was obtained with ISAAC at the ESO VLT-UT1 
({\it Antu}) 
on 25-27 April 1999 in photometric and 0.5$^{\prime\prime}$-1.0$^{\prime\prime
}$ seeing conditions. The slit was 1$^{\prime\prime}$ wide, providing 
a spectral resolution FWHM of $\sim$ 24~\AA, 32~\AA~ and 48~\AA~ in $J$-,
$H$-, and $K$-band 
respectively. The observations were done by nodding the target along the 
slit between two positions A and B with a nod throw of 10$^{\prime\prime}$.
The integration time per position was 10 minutes. For instance, a total 
integration time of 1 hour was obtained following a pattern ABBAAB.
When possible, two targets were observed simultaneously in the slit. 
The spectral frames were flat-fielded, 
rectified, sky-subtracted, coadded and divided by the response curve 
obtained using the spectra of O7-O8 stars. The spectra were extracted 
using a 8 pixel wide aperture. Absolute flux calibration was achieved 
by normalizing the spectra to the $JHKs$ SOFI or ISAAC broad-band
photometry. Table 1 lists the main information about the targets and 
their observation.

\begin{table*}
\caption[]{The observed sample}
\begin{tabular}{llrlllllll} \hline\hline
 & & & & & & & & & \\
Target & $K(3^{\prime \prime})$ & Color & Spectra & Integration &
$z_{{\rm sphot}}$ & $F_{{\rm lim}}$ & $SFR$ & Morphology & Class \\
       &      &       &     & (hours) & & & ($h_{50}^{-2}$M$_{\odot}$yr$^{-1}$) & \\
J100551-0742.4 & 18.6 & $I-K$=4.8 & $JHK$ & 1,1,1 & 1.80$\pm$0.10 & $<$1.8 & $<31$ & c,r & E \\
J100544-0742.2 & 18.4 & $I-K$=4.6 & $JHK$ & 1,1,1 & 0.80$\pm$0.30 & $<$4.5 &$<13$ & e,i & D \\
J101948-2219.8 & 18.7 & $R-K$=6.6 & $H$ & 1 & 1.52$\pm$0.12 & $<$1.8 & $<22$ & & E \\
J101950-2220.9 & 18.6 & $R-K$=6.9 & $H$ & 1 & 1.50$\pm$0.25 & $<$1.8 & $<23$ & & E \\
J124027-1131.0 & 18.2 & $I-K$=4.6 & $JHK$ & 2,1,2 & 0.90$\pm$0.30 & $<$1.7 & $<7$ & c,wd & D \\
J202759-2140.8 & 17.9 & $R-K$=5.1 & $JHK$ & 1,2.2,1 &0.80$\pm$0.12 & $<$2.0 &$<7$ & & E \\
J202800-2140.9 & 18.1 & $R-K$=5.1 & $JH$ & 1,2.2 & 0.82$\pm$0.20& $<$2.0 & $<6$ & & E \\
J202807-2141.1 & 17.7 & $R-K$=5.9 & $HK$ & 1.3,1.3 & 0.88$\pm$0.15 & --- & --- & & D? \\
J202807-2140.8 & 18.4 & $R-K$=5.9 & $HK$ & 1.3,1.3 & 0.96$\pm$0.15 & --- & --- & & E \\
 & & & & & & & & \\\hline\hline
\end{tabular}

\small
Notes:

(1) $F_{{\rm lim}}$: emission line limiting flux in units of $10^{-16}$ erg 
s$^{-1}$ cm$^{-2}$. 

(2) The field J1019-223 includes a radiogalaxy at $z=1.77$ (McCarthy et 
al. 1996). 

(3) The field J2027-217 includes a radiogalaxy at $z=2.63$ (McCarthy et 
al. 1992).

(4) J202759-2140.8 and J202800-2140.9 were called respectively ``a'' and ``c'' 
by McCarthy et al. (1992). 

(5) Morphology in $I_{{\rm 814}}$ band: c=compact, e=extended, r=regular, 
i=irregular, wd= weakly disturbed. 

(6) Class: classification based on SED fitting: E=elliptical, D=dusty 
(see text for more details).

\normalsize

\end{table*}

\section{Results}

\subsection{ISAAC spectroscopy}

The main aim of ISAAC spectroscopy was to check for the presence of
redshifted 
emission lines such as H$\alpha$ at $z>0.7$, [OIII]$\lambda$5007 at 
$z>1.2$ and [OII]$\lambda$3727 at $z>2.0$ in order to assess what 
fraction of ERGs is made by objects such as starburst or active galaxies. 
For four targets we could cover the whole $JHK$ spectral range. 

Continuum emission was detected in all the targets. However, neither 
strong emission lines nor evident continuum breaks were detected 
in the ISAAC spectra. The absence of emission lines favours these ERGs being 
high-$z$ ellipticals, although this does not prove it. In case of elliptical 
galaxies, the absence of the 4000~\AA~ continuum break in the observed
$J$-band spectra contrains the redshifts to be at $z<1.8$. The possibility 
that H$\alpha$ emission falls in the ``blind'' region between $J$ and $H$ bands 
(for 1.1$< z<$1.2) with [OIII]$\lambda$5007 not yet entered in the $J$ band, 
seems unlikely to occur for all the targets.

The upper limits for the flux of emission lines are rather similar for 
all the spectra. However, each spectrum shows a variation of this limit 
as a function of wavelength because of the enhanced 
noise in regions corresponding to bright sky emission lines. The deepest
limits are $F_{{\rm lim}}<$1.0$\times10^{-16}$ erg s$^{-1}$ cm$^{-2}$ and 
$W_{{\rm lim}}<$
50~\AA, increasing by a factor of $\approx 2$ in the parts of the
spectra most contaminated by sky line residuals. Due to its fainter and noisier spectra, the exception is 
J100544-0742.2 which has $F_{{\rm lim}}<$4.5$\times10^{-16}$ erg 
s$^{-1}$cm$^{-2}$ 
and $F_{{\rm lim}}<$3.5$\times10^{-16}$ erg s$^{-1}$cm$^{-2}$ in $J$ and $H$, 
respectively. With the exception of J100544-0742.2, our spectra 
would allow to significantly detect emission lines such as the H$\alpha$ 
observed in HR10 (Dey et al. 1999), or in a major fraction of the 
H$\alpha$ emitters at $1.1<z<1.9$ discovered by McCarthy et al. (1999). 

\subsection{Spectro-photometric redshifts}

In absence of distinctive spectral features, we attempted to estimate 
what we call the ``spectro-photometric'' redshifts ($z_{{\rm sphot}}$) by using 
the ISAAC continuum spectrophotometry together with the available broad-
band photometry in order to derive the global optical to near-IR spectral
energy distributions (SEDs) of the observed ERGs. The resulting SEDs 
were then fitted by means of a $\chi^{2}$ comparison with a set of 
240 Bruzual \& Charlot (1998) synthetic templates: 80 simple stellar 
population (SSP) spectra with $Z=Z_{\odot}$, 80 with constant star 
formation (CSF) and $Z=Z_{\odot}$, and 80 SSP spectra with $Z=2.5 
Z_{\odot}$. The template spectra have Salpeter IMF (0.1$<$M$_{\odot}<$
125) and ages from 0.1 Myr to 10 Gyr. Such a set of templates was chosen
in order to encompass a wide range of galaxy SEDs spanning from 
young starbursts to old ellipticals with super-solar metallicity.
The dust extinction was 
treated as a free parameter adopting the Calzetti (1997) extinction 
law with $E_{{\rm B-V}}$=0.0-1.0. The $J$ and $H$ spectra of J100544-0742.2 
were not used in the fitting analysis because too noisy. 

\begin{figure*}
\resizebox{16cm}{!}{\includegraphics{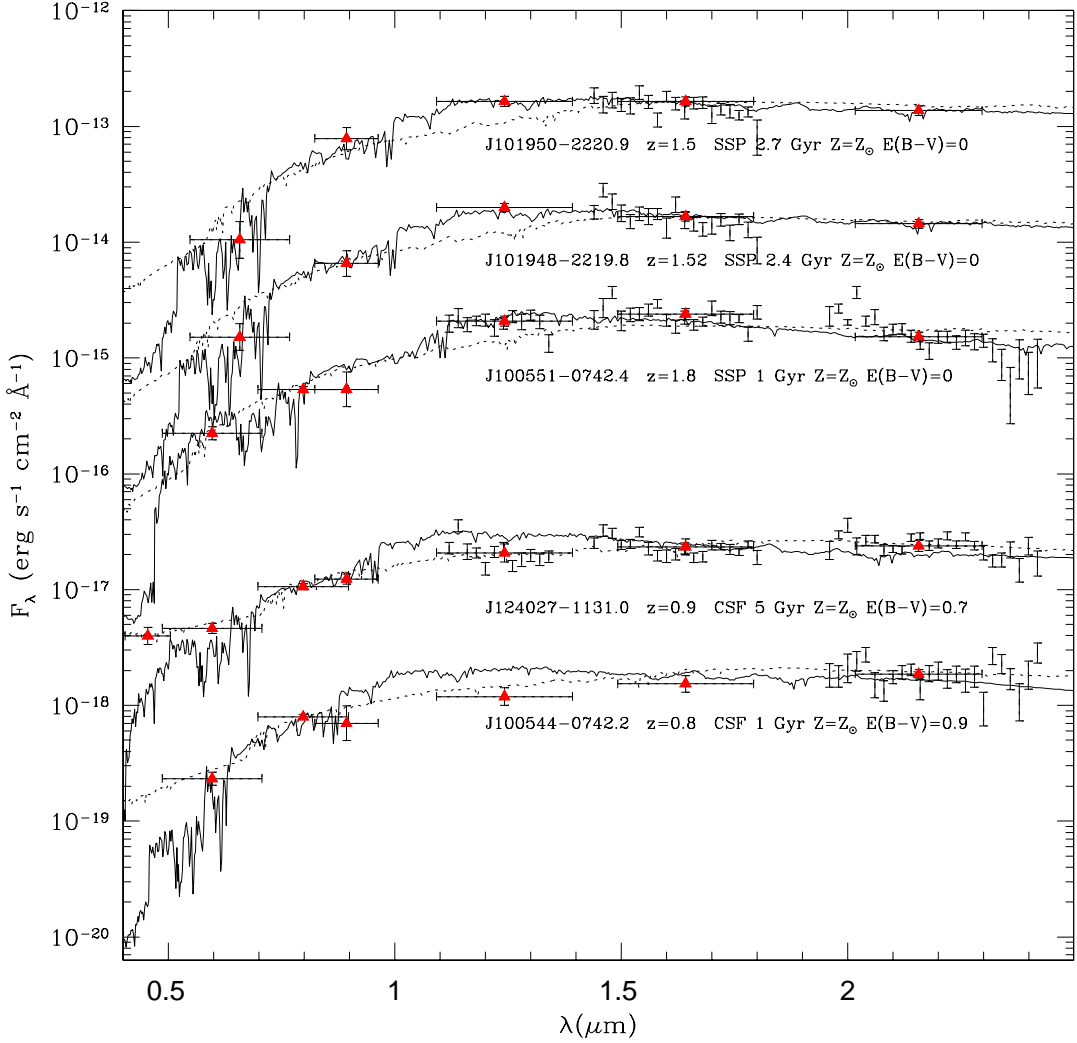}}
\hfill
\caption{
The global optical to near-IR SEDs based on broad-band photometry 
(filled triangles) and ISAAC spectroscopy (vertical bars). The
ISAAC spectra shown are binned to 200~\AA. The first three
SEDs from the top are those consistent with no dust extinction. The 
continuum and the dotted lines are the best fits obtained respectively 
with dustless SSP spectra and with dusty CSF spectra ($Z=Z_{\odot}$). 
Such fits are shown in order to provide a comparison between the dusty 
and the dustless cases. The best fitting parameters ($z$, ages and 
$E_{{\rm B-V}}$) are shown relatively to the best fit of the two cases 
(i.e. dusty or dustless). Optical photometry was obtained with the 
following telescopes, instruments and filters: ESO NTT+SUSI ($R$) for 
J101950-2220.9 and J101948-2219.8, HST+WFPC2 ($V_{{\rm 606}}, I_{{\rm 
814}}$) for J100544-0742.2 and J100551-0742.4, and HST+WFPC2 ($B_{{\rm 
450}}, V_{{\rm 606}}, I_{{\rm 814}}$) for J124027-1131.0. From bottom 
to top, the five SEDs are multiplied by $10^{0,1,3,4,5}$ respectively.
}
\end{figure*}

During the fitting analysis, the ISAAC $JHK$ spectra were first 
cleaned by excluding the edge regions and, in the $H$-band, the 
regions corresponding to the four strongest sky emission lines. 
The ISAAC spectra were then binned to 800~\AA~ wide bins in order to
balance the relative contribution of spectrophotometry and 
broad-band photometry during the fitting analysis, and, at 
the same time, to conserve the spectral information (i.e. the 
slope of the spectra). Broad-band $JHK_{\rm s}$ photometric points were 
not used during the fitting if corresponding near-IR spectra were 
available. The $z_{{\rm sphot}}$ shown in Tab. 1 are relative to the
best absolute fit. The formal 1$\sigma$ uncertainties on the redshifts, 
estimated with the $\Delta \chi^{2}$ method (Avni 1976), are 
$\sigma_{{\rm z_{sphot}}} \sim 0.1-0.3$ (see Tab.1). Figures 1 and 2 show 
that a good agreement between the observed and the model spectra 
is generally achieved. In the case of J100551-0742.4, the absence 
of an observed 4000~\AA~ break in the $J$-band spectrum constrains 
$z_{{\rm sphot}}<1.8$, whereas the broad-band photometry would allow 
to derive formally acceptable redshifts up to $z>2$. 

\subsection{Dustless and dusty ERGs}

According to the results of the fitting analysis, we attempted to
preliminarly divide the observed ERGs into two classes at a 
95\% confidence level. 

{\it (A)} Galaxies with SEDs that can be reproduced by evolved
stellar population spectra without dust extinction. 6 ERGs fall 
into this class (see Fig. 1 and 2). This does not necessarily mean 
that these galaxies are dustless, but it only means that their SEDs 
are consistent with those expected for passively evolved high-$z$ 
ellipticals with no dust extinction, old ages ($\sim 1-3$ Gyr) and 
metallicity up to $Z=2.5Z_{\odot}$. However, because of the degeneracy 
between ages and dust extinction, acceptable fits can be found also 
with younger stellar populations and $0<E_{{\rm B-V}}<0.4$. Thus, we can 
summarize saying that {\it at most} 6 ERGs of our sample can be 
high-$z$ ellipticals. In such a case, their rest-frame $K$-band 
absolute magnitudes ($M_{{\rm K}}\sim$-24.4$\div$-25.0) imply luminosities 
$L \lesssim L^{\ast}$ (adopting $M_{{\rm K}}^{\ast}$=-25.16 for the local 
luminosity function of elliptical galaxies; Marzke et al. 1998), and 
their stellar masses are intermediate (M$_{\ast} \sim 1-4\times 
10^{11} h_{50}^{-2}$ M$_{\odot}$). 

\begin{figure*}
\resizebox{16cm}{!}{\includegraphics{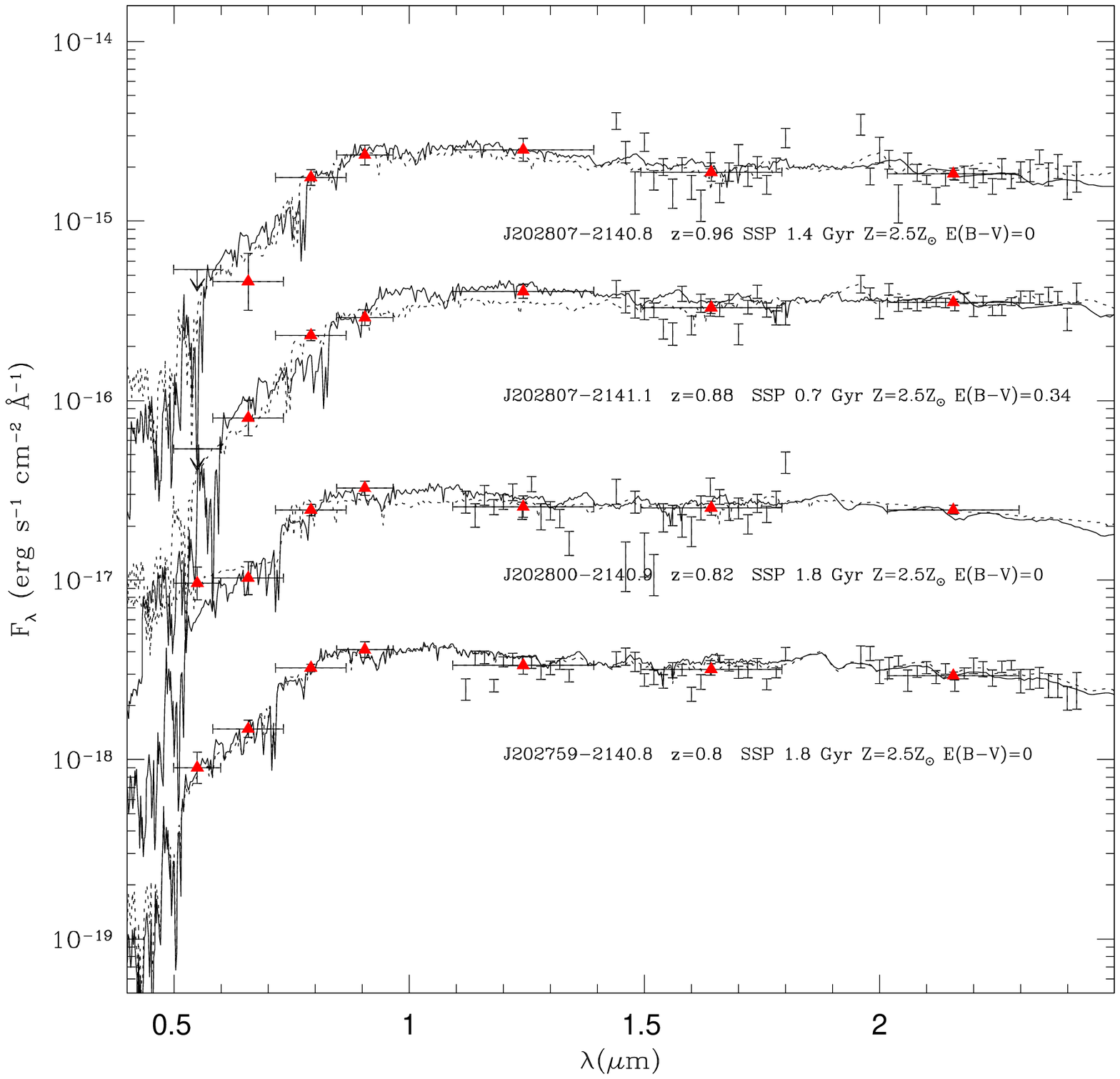}}
\hfill
\caption{
The global optical to near-IR SEDs (same symbols as 
in Fig. 1). Here the continuum and the dotted lines are the best fits 
obtained respectively with dustless and dusty SSP spectra (all with 
$Z=2.5Z_{\odot}$). Optical photometry comes from ESO NTT+EMMI observations 
with $VRIZ$ filters. From bottom to top, the four SEDs are multiplied by 
$10^{0,1,2,3}$ respectively.
}
\end{figure*}

{\it (B)} Galaxies with SEDs that require significant dust extinction.
Two ERGs are present in this group: J100544-0742.2 and J124027-1131.0.
Their SED fitting is characterized by a large degree of degeneracy 
among spectral templates, ages and amount of dust reddening. Nevertheless,
the minimum amount of dust extinction required to provide acceptable fits 
($E_{{\rm B-V}}>0.5$) suggests that these two ERGs are systems affected by 
strong dust reddening. A third galaxy, J202807-2141.1, falls formally
into this class because an acceptable fit is obtained only with 
$E_{{\rm B-V}}\gtrsim 0.3$, with an age of $\approx$0.7 Gyr and with
$Z=2.5Z_{\odot}$. However, we consider this case ambiguous because 
of the unusual set of parameters provided by the fit. Deep submillimeter 
continuum observations will help to establish if these three dusty ERGs 
are HR10-like objects (e.g. Cimatti et al. 1998).

HST WFPC2 deep imaging is available from the public archive
only for three of the observed ERGs (see Tab. 1). The 
$I_{{\rm 814}}$-band morphologies strengthen the indications given by 
the SED fitting analysis: J100544-0742.2 has an extended, irregular 
and disturbed morphology consistent with being a dusty starburst galaxy; 
J100551-0742.4 has a compact and elliptical-like morphology. The case of 
J124027-1131.0 is more ambiguous because its morphology is compact 
and weakly disturbed. 

Adopting $z=z_{{\rm sphot}}$, for each object we estimated the limits on the
H$\alpha$ luminosity $L($H$\alpha)$ and on the star formation rate ($SFR$) 
through the relation $SFR=7.9\times10^{-42}L($H$\alpha)$ [M$_{\odot}$
yr$^{-1}$] (Kennicutt 1998). This was not possible for J202807-2141.1 and 
J202807-2140.8 because no $J$ spectra were available to us. We find 
limits in the range of $L($H$\alpha)< 7-40 \times 10^{41}h_{50}^{-2}$ 
erg s$^{-1}$, corresponding to $SFR < 6-30$$h_{50}^{-2} $M$_{\odot}$
yr$^{-1}$ (Table 1). Taken at face value, these limits imply star
formation rates at most as high as those of nearby gas rich spiral 
galaxies (see Kennicutt 1998). However, in case of dust extinction, such 
SFRs would increase by a factor of $\approx$2--30$\times$ for $E_{{\rm B-V}}=$
0.2--0.9.

Finally, we noticed that the four ERGs in the J2027-217 field have 
all very similar $z_{{\rm sphot}}$ (Fig.2), and may belong to a same cluster
or group. 

\section{Summary and main implications}

Our observations showed that neither strong emission lines nor continuum
breaks are detected in the ISAAC spectra of 9 ERGs, and
that only a fraction of them (2-3 out of 9 in our subsample) require strong
dust reddening to reproduce their SEDs. On the other hand, up to 6 ERGs 
have properties that are formally consistent with the strict definition of 
being dustless, old and passively evolved spheroidals at $z \gtrsim 1$
and with $z_{{\rm formation}}>2$. 
Although the observed sample is rather small and still incomplete, 
it is tempting to speculate on how the above results may have
implications on the problem of the formation of elliptical galaxies. 

In fact, the existence and the abundance of high-$z$ ellipticals
is one of the most controversial issues of galaxy evolution.
Some works claimed that the number of galaxies with the red colors 
expected for high-$z$ passively evolved spheroidals is lower 
compared to the predictions of passive luminosity evolution (e.g. 
Kauffmann, Charlot \& White 1996; Zepf 1997; Franceschini et al. 1998; 
Barger et al. 1999). However, other works did not confirm the existence of 
such a deficit up to $z\approx2$ (e.g. Totani \& Yoshii 1997; Benitez 
et al. 1999; Broadhurst \& Bowens 1999; Schade et al. 1999). This  
picture is complicated by the possibility that such spheroidals are 
bluer because of a low level of residual star formation, thus escaping 
the selection criteria based on red colors (e.g. Jimenez et al. 1999). 

Thus, it is clear that a reliable comparison between the observed
abundance of high-$z$ ellipticals and the one expected from 
the various galaxy formation models can be performed only if the 
fraction of high-$z$ ellipticals in ERG samples is firmly established. 
If taken at face value, our analysis suggests that a substantial fraction 
of ERGs are consistent with being {\it bona fide} high-$z$ spheroidals. 
Should spectroscopy of complete and larger samples confirm such 
result, this would allow a reliable comparison of the observed and 
predicted numbers of spheroidals at $z \gtrsim 1$. 
%
%
%
%
%For instance, we can compare the observed surface density of ERGs 
%with $R-K>5$ and $K^{\prime}<19$ ($\sim 0.43 \pm 0.05$ arcmin$^{-2}$);
%Thompson et al. 1999) with the expected abundance of ellipticals
%
%
%
%
%
\begin{acknowledgements}
We are grateful to Jean-Gabriel Cuby, Chris Lidman and Leonardo Vanzi for 
their invaluable help during the observations, and to Gustavo Bruzual and
Stephane Charlot for providing their synthetic spectral library. We also
thank the referee, Massimo Stiavelli, for his constructive criticism.
\end{acknowledgements}

{}

\end{document}